\documentclass[twocolumn,floatfix,pre,aps,showpacs]{revtex4}
\input{graphicx}
\def \ind#1{{\mbox{\footnotesize {#1}}}}
\def \ind#1{{\mbox{\scriptsize {#1}}}}

\begin{document}

\title{Self-Organized Criticality Below The Glass Transition}
\author{Katharina Vollmayr-Lee$^{1,2}$ and Elizabeth A. Baker$^{1}$}
\affiliation{$^{1}$ Department of Physics and Astronomy, Bucknell University,  
      Lewisburg, Pennsylvania 17837, USA\\
   $^{2}$ Institut f\"ur Theoretische Physik, Universit\"at G\"ottingen,
       Friedrich-Hund-Platz 1, D-37077 G\"ottingen, Germany}
\email{kvollmay@bucknell.edu}
\date{March 26, 2006}

\begin{abstract}
We obtain evidence that the dynamics of glassy systems below the glass
transition is characterized by self-organized criticality.  Using
molecular dynamics simulations of a model glass-former we identify
clusters of cooperatively jumping particles.  We find string-like
clusters whose size is power-law distributed not only close to
$T_\ind{c}$ but for {\em all} temperatures below $T_\ind{c}$,
indicating self-organized criticality which we interpret as a freezing
in of critical behavior.
\end{abstract}

\maketitle

Although the relaxation dynamics of glass-forming liquids has been
studied for decades, there are still many unresolved questions
\cite{ref:Binder_Kob}. Especially for the dynamics of the glass out of
equilibrium it is still an open and hotly debated question what
characterizes the relaxation.  We present in this Letter work on the
dynamics below the glass transition where we focus on cooperative
motion.  Cooperative rearranging regions have been studied mostly
above the glass transition and are the basis of Adam-Gibbs theory
\cite{ref:Adam_Gibbs}.  Above the glass transition two kinds of
cooperative motion have been identified: (i) string-like motion
\cite{ref:string_Donati,ref:Weeks_Science2000,ref:string_other} where
of the order of ten particles move along a Conga-line and where each
particle is significantly more mobile than an average particle, and
(ii) very cooperative motion where of the order of 40 particles
participate and where each particle undergoes only a small
displacement \cite{ref:appignanesi}.

To study cooperative motion below the glass transition we use
molecular dynamics simulations, which have the advantage of providing
us with the microscopic information of every particle's position at
all times. Using these particles trajectories we first search 
each simulation run for jump events where a particle jumps out of its
cage of neighbors.  Then we identify clusters of cooperatively jumping
particles, i.e.  jump events which are correlated in space and time.
We find that the cluster size distribution follows a power law
independent of details of the cluster definition.  Furthermore, we
find string-like clusters as they have been found above the glass
transition.

\begin{figure}[htbp]
\includegraphics[width=3.4in]{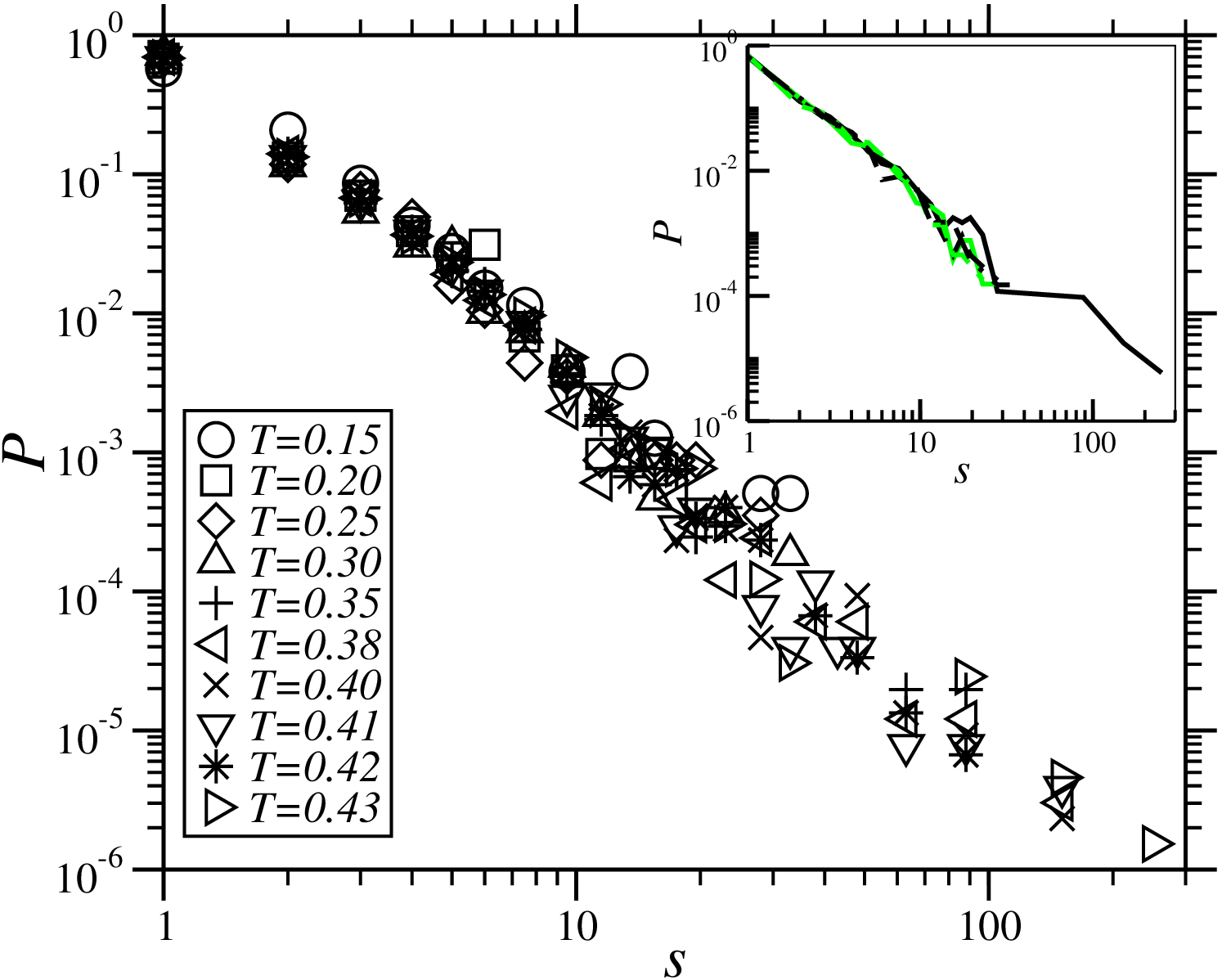}
\caption{\sf Distribution $P(s)$ of cluster sizes of simultaneously
  jumping particles for temperatures $T=0.15 - 0.43$.  The inset shows
  $P(s)$ at $T=0.43$ where the jump events have been divided into five
  equal time windows.  The black solid line represents the first
  window. }
\label{fig:fig1}
\end{figure}

A similar cluster definition for the system under study but at a
temperature slightly above the glass transition has also revealed a power
law distribution \cite{ref:string_Donati}.  Such distributions are a
signature of criticality, such as the cluster size distribution in
percolation theory at the critical point
\cite{ref:Stauffer_review}. However, contrary to these simulations and
percolation theory, we find a power law not only close to a critical
point but for all temperatures below $T_\ind{c}$ that we have
investigated (see Fig.~\ref{fig:fig1}).  We thus find a type of
self-organized criticality.

Our data are consistent with the following scenario: a glass cooled
down to $T_\ind{c}$ develops critical behavior, and then upon further
cooling the criticality remains frozen in.  Mode-coupling theory for
glasses predicts the development of critical fluctuations
\cite{ref:Goetze} when $T_\ind{c}$ is approached from above and a
recent extension of mode-coupling theory to temperatures below the
transition has predicted such a freezing in of critical behavior
\cite{ref:Latz_condmat}. The signature of criticality in these
theories is the power law of a two-time correlation function
\cite{ref:Latz_condmat,ref:Latz2000,ref:Bouchaud_etal}. Our data,
while consistent with the mode-coupling freezing in scenario, find
criticality in a spatial structure rather than a relaxation
exponent. We believe this represents the first direct observation in
glasses of self-organized criticality, that is, criticality for all
temperatures below $T_\ind{c}$.

Our system is a well-studied binary Lennard-Jones (LJ) mixture of 800
A and 200 B particles.  We refer the reader for details of the model
to \cite{ref:kob_andersen} and for details of the molecular dynamics
simulations to \cite{ref:JCP2004}.  Previous simulations have shown
that this system exhibits the main features of glass-forming liquids
and is thus a good simple model for glass-formers
\cite{ref:kob_andersen}.  For present-day computer simulations the
system falls out of equilibrium in the vicinity of $T_\ind{g} = 0.435$
(in reduced LJ units) \cite{ref:kob_andersen}.  Whereas Donati {\it et
al.}  \cite{ref:string_Donati} studied cooperative motion of this
system {\em above} the glass transition, we study here the same system
but {\em below} the glass transition at 10 temperatures ranging from
$0.15$ to $0.43$.  We use 10 independent, well-equilibrated
configurations at $T=0.5$ and then instantly quench the system to the
desired temperature, e.g. $T=0.15$. After an (NVT) run of 2000 time
units we then run the (NVE) production run for $10^5$ time units.

For the definition of jump events we use the trajectory ${\bf r}_n(t)$
of each particle $n$ and take time averages over 800 time units to
obtain its thermal fluctuation $\sigma_n$ and average positions
$\overline{{\bf r}_n}(t_l)$ at times $t_l = 800 (l-0.5)$ where $l=1,
2, \ldots, 125$.  We define a particle $n$ to undergo a jump if its
change in average position $\Delta \overline{r_n}= \left |
\overline{{\bf r}_n}(t_l) - \overline{{\bf r}_n}(t_{l-4}) \right |$
satisfies $\Delta \overline{r_n} > \sqrt{20} \, \sigma_n$
\cite{ref:sigestimate}.  We thus identify for the whole simulation run
all jump events $\{n, l_i, \langle {\bf r}_n \rangle_i, l_f, \langle
{\bf r}_n \rangle_f\}$ of jumping particles $n$, jumping from average
position $\langle {\bf r}_n \rangle_i$ at time $t_{l_i}$, the time
associated with bin $l_i$, to average position $\langle {\bf r}_n
\rangle_f$ at time $t_{l_f}$ \cite{ref:more_details}.  

To address the question of cooperative motion we investigate how these
single particle jump events are correlated in time and space.  To
identify correlations in time, we group the jump events according to
the bin index $l_i$.  We thus obtain $N_l$ simultaneously jumping
particles for each time bin $l$.  To investigate how these $N_l$
particles are spatially correlated, we identify clusters where
particles $n$ and $m$ are defined to be neighbors (and therefore
members of the same cluster) if their distance $|\langle {\bf r}_n
\rangle_i-\langle {\bf r}_m \rangle_i|$ is smaller than the position
of the first minimum $r_\ind{min}$ of the corresponding radial pair
distribution function of the complete system ($r_\ind{min}=1.4$ for
AA, $1.2$ for AB and $1.07$ for BB independent of temperature).  This
analysis gives us for each time $K_l \ge 0$ distinct clusters. The
clusters are numbered by $k=1, 2, \ldots, K_l$ and we denote by ${\cal
N}_{l,k}$ the set of particle labels composing the $k$th cluster in
time bin $l$ with $N_{l,k}$ particles (i.e. $\sum \limits_{k=1}^{K_l}
N_{l,k} = N_l$).  We now look at the size distribution $P(s)$ of all
clusters ${\cal N}_{l,k}$, with $s$ being the number of cluster
members, i.e.
\begin{equation}
\label{eq:P(s)}
P(s) = \sum \limits_{l} \, \sum \limits_{k=1}^{K_l} \,
                    \delta (s,N_{l,k})
              \bigg /
            \sum \limits_{l} \, K_l
\end{equation}
where $\delta(x,y)$ is the Kronecker delta function.

In Fig.~\ref{fig:fig1} we show a log-log plot of $P(s)$ for
temperatures $T=0.15-0.43$. We observe essentially a straight line,
i.e. a power law $P(s) \sim s^{-\tau}$, indicating scale invariance.
The exponent for different temperatures ranges from $\tau=2.2-2.7$
with the trend of $\tau$ increasing with temperature.  The average is
$\tau=2.50\pm0.05$. A power law of $P(s)$ has also been found in
simulations of the same binary Lennard-Jones system slightly above but
close to $T_\ind{c}$ with $\tau=1.86$ \cite{ref:string_Donati}. In
percolation theory the size distribution $n_s$ (where $P(s)=n_s /
\sum_s n_s$) follows also a power law where the mean-field exponent
$\tau=2.5$ and in three dimensions $\tau=2.2$
\cite{ref:Stauffer_review}.  However, in these simulations and in
percolation theory the power law occurs only {\em at} $T_\ind{c}$. In
contrast, we find a power law for all temperatures (see
Fig.~\ref{fig:fig1}), i.e. we find scaling invariance for the {\em
whole} temperature range below $T_\ind{c}$.  As mentioned above, this
is consistent with the scenario of critical behavior being frozen in,
and remaining for all temperatures below $T_\ind{c}$.

This scale-invariance independent of a control parameter is usually
found in systems with self-organized criticality
\cite{ref:selforgan_crit} and is, to our knowledge, a new phenomenon
for the cluster size distribution of structural glass formers.  The
occurrence of self-organized criticality is usually associated with
being out of equilibrium and having widely separated time scales.  Our
system is consistent with these requirements: single particle jumps
take of the order of 10 time units, the time between successive jumps
is of the order of 30000 \cite{ref:JCP2004}, and the equilibrium
relaxation time is significantly longer than the simulation run
\cite{ref:kob_andersen}.

Due to the importance of implications for relaxation dynamics below
the glass transition, the question arises if our results for the
cluster size distribution are specific to details of the analysis.  We
therefore investigate how robust $P(s)$ is with
respect to different definitions.  Since the system is out of
equilibrium, we first ask the question if the jump dynamics changes
during the simulation run. We therefore divide the whole simulation
run into five time windows of equal size and determine $P(s)$ with
Eq.(\ref{eq:P(s)}) by restricting the sum over $l$ accordingly.  The
inset of Fig.~\ref{fig:fig1} shows the resulting $P(s)$ for
$T=0.43$. We find that in the first time window the most collective
processes occur: up to approximately 250 particles jump simultaneously
and spatially correlated. The distribution $P(s)$ however seems to
follow a power law with $\tau=2.5\pm0.1$ not only independent of
temperature but also independent of the time window (i.e.  waiting
time).  We find the same waiting time independence also for all other
temperatures.

\begin{figure}[htbp]
\includegraphics[width=3.4in]{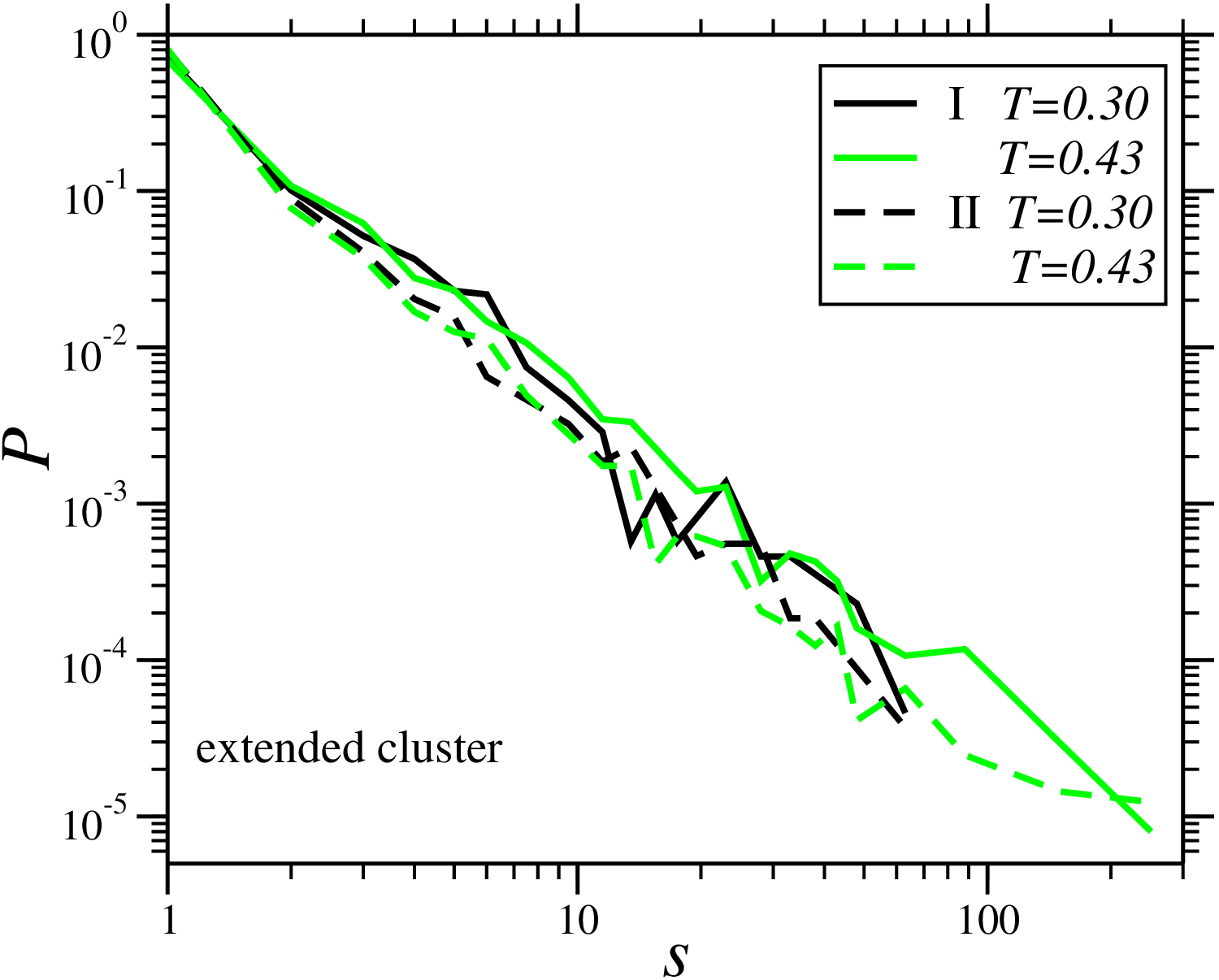}
\caption{\sf 
  Distribution of cluster size for extended clusters I \& II for 
temperatures $T=0.30$ and $0.43$.
}
\label{fig:fig2}
\end{figure}

To further check the sensitivity on details of the power law of $P(s)$
we next modify the definition of a cluster.  Whereas usually cluster
connections are defined in space (and therefore avalanche-like
correlations are tested in space), we generalize now the definition of
a cluster by treating space and time similarly (and thus allowing for
avalanche-like correlations in time also).  Instead of requiring as
before that two jump events $\alpha$ and $\beta$ occur simultaneously
($l_i^{\alpha} = l_i^{\beta}$) we define extended clusters by allowing
two jump events to occur at neighboring time bins (i.e. $|\Delta l| \le
1$).  We show in Fig.~\ref{fig:fig2} results for two
different definitions of extended clusters (I \& II). The difference
between these definitions is due to the usage of time and position
before or after the jump.  We define two jump events $\{n^{\alpha},
l_i^{\alpha}, \langle {\bf r}_n \rangle_i^{\alpha}, l_f^{\alpha},
\langle {\bf r}_n \rangle_f^{\alpha}\}$ and $\{m^{\beta}, l_i^{\beta},
\langle {\bf r}_m \rangle_i^{\beta}, l_f^{\beta}, \langle {\bf r}_m
\rangle_f^{\beta}\}$ to be connected if
\begin{tabular}{llclr}
def.I: & $|l_i^{\alpha} - l_i^{\beta}| \le 1$ & and & 
         $|\langle {\bf r}_n \rangle_i^{\alpha} 
         - \langle {\bf r}_m \rangle_i^{\beta}| 
                               \le r_\ind{min}$ & \\
def.II: & $\big ( |l_i^{\alpha} - l_f^{\beta}| \le 1$ & and & 
          $|\langle {\bf r}_n \rangle_i^{\alpha} 
             - \langle {\bf r}_m \rangle_f^{\beta}|  
                               \le r_\ind{min} \big )$  &  or\\
        & $\big ( |l_f^{\alpha} - l_i^{\beta}| \le 1$ & and & 
          $|\langle {\bf r}_n \rangle_f^{\alpha} 
              - \langle {\bf r}_m \rangle_i^{\beta}|  
                               \le r_\ind{min} \big )$  &  \\
\end{tabular}

As before for simultaneously jumping particles, definitions I and II
result in a cluster size distribution which follows a power law,
however, with a smaller exponent of $\tau = 2.13\pm0.04$ and
$\tau=2.22\pm0.04$ for clusters I and II respectively. As shown in
Fig.~\ref{fig:fig2} for $T=0.30$ and $0.43$ we find again a power law
for all temperatures (similar results are obtained for other
temperatures).

\begin{figure}[t]
\includegraphics[width=3.4in]{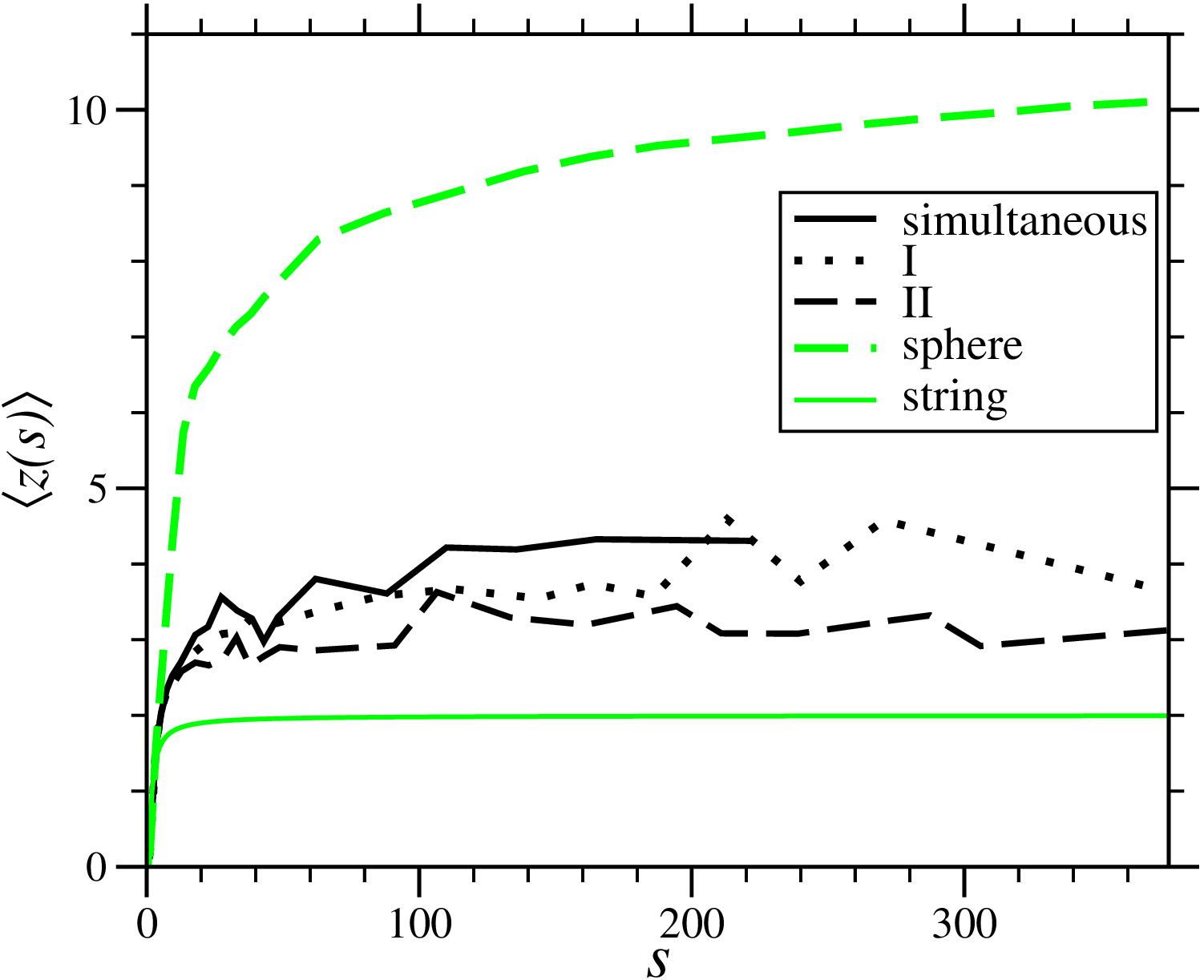}
\caption{\sf 
  Average coordination number $z$ within a cluster as a function of 
  cluster size $s$. Error bars are of the order of 
$\Delta \langle z \rangle =0.5$.
}
\label{fig:zofs}
\end{figure}

To further investigate these clusters we next characterize their
geometric shape directly via coordination numbers (instead of 
displacement-displacement correlations as in \cite{ref:string_Donati}).
We determine for each
cluster ${\cal N}_{l,k}$ the average coordination number 
\begin{equation}
\label{eq:z(s)}
z_{l,k} = \frac{1}{N_{l,k}} \, 
       \sum \limits_{n \in {\cal N}_{l,k}} z_n
\end{equation}
where $z_n$ is the number of neighboring particles $m \in {\cal N}_{l,k}$
of particle $n$.
Fig.~\ref{fig:zofs} shows the average 
\begin{equation}
\langle z(s) \rangle = 
           \sum \limits_{l} \, \sum \limits_{k=1}^{K_l} \,
                       \delta (s,N_{l,k}) z_{l,k}
              \bigg /
            \sum \limits_{l} \, \sum \limits_{k=1}^{K_l} \,
                       \delta (s,N_{l,k})
\end{equation}
as a function of $s$. We observe no temperature dependence of 
$\langle z(s) \rangle$ and therefore an
additional average over simulation runs at different temperatures has 
been included in Fig.~\ref{fig:zofs}.
The comparison with an ideal string and the most compact cluster
(sphere) indicates that both the clusters of simultaneously jumping 
particles as well as the extended clusters are string-like. This is
similar to the results of cooperative motion above the glass transition 
\cite{ref:string_Donati,ref:Weeks_Science2000,ref:string_other}
and below the glass transition \cite{ref:coop_belowTg}.

Our results are consistent with the following scenario:
above the critical temperature $T_\ind{c}$ string-like clusters 
are found.
Close to $T_\ind{c}$ the distribution of cluster sizes follows a power
law. Below the glass transition this critical behavior gets frozen in.
Independent of details of the cluster definition and independent of 
waiting time,
we find string-like clusters
with a cluster size distribution
which follows a power law for all investigated temperatures.
We expect this self-organized criticality 
to occur also for other
glasses out of equilibrium and leave the test thereof for future work.

KVL thanks the Institute of Theoretical Physics, University
G\"ottingen, for hospitality and financial 
support. 
EAB gratefully acknowledges support from NSF Grant No. REU-0097424.
The authors thank J. Horbach, W. Kob and K. Binder for comments on
an earlier version of this manuscript and also
A. Latz, and A. Zippelius for helpful discussions.


\end{document}